\documentclass[journal]{IEEEtran}
\usepackage[utf8]{inputenc}
\usepackage{amsmath}
\usepackage{graphicx}
\usepackage{fancyhdr}
\usepackage{booktabs}
\usepackage{multirow}
\usepackage{subfig}
\usepackage{balance}
\usepackage[]{url}
\usepackage{color}
\begin{document}
\title{The Influence of Social Ties on Performance in Team-based Online Games}

\author{Yilei~Zeng,~\IEEEmembership{Student Member,~IEEE,}
        Anna~Sapienza,~\IEEEmembership{Member,~IEEE,}
        and~Emilio~Ferrara,~\IEEEmembership{Senior~Member,~IEEE}% <-this % stops a space
\thanks{USC Information Sciences Institute, Marina del Rey, CA 90292 (USA)}% <-this % stops a space
\thanks{USC Department of Computer Science, Los Angeles, CA 90089 (USA)}% <-this % stops a space
\thanks{Manuscript received ...; revised ...}}
%\markboth{Journal of \LaTeX\ Class Files,~Vol.~14, No.~8, Nov~2018}%
\maketitle

\begin{abstract}
Social ties are the invisible glue that keeps together human ecosystems. Despite the massive amount of research studying the role of social ties in communities (groups, teams, etc.) and society at large, little attention has been devoted to study their interplay with other human behavioral dynamics. 
Of particular interest is the influence that social ties have on human performance in collaborative team-based settings.
Our research aims to elucidate the influence of social ties on individual and team performance dynamics. We will focus on a popular \textit{Multiplayer Online Battle Arena} (MOBA) collaborative team-based game, \textit{Defense of the Ancients 2} (Dota 2), a rich dataset with millions of players and matches.
Our research reveals that, when playing with their friends, individuals are systematically more active in the game as opposed to taking part in a team of strangers. However, we find that increased activity does not homogeneously lead to an improvement in players’ performance. Despite being beneficial to low skill players, playing with friends negatively affects performance of high skill players. 
Our findings shed light on the mixed influence of social ties on performance, and can inform new perspectives on virtual team management and on behavioral incentives.
\end{abstract}

\IEEEpeerreviewmaketitle

\section{Introduction}
\IEEEPARstart{T}{he} interplay between social networks and complex human behavioral dynamics has been the subject of extensive research. Social ties, a.k.a. friendships and/or acquaintanceships, have been connected to various  complex human phenomena. Some studies, for example, demonstrated the social capital brought about by strong  and weak ties~\cite{granovetter1977strength}. Others illuminated on the benefits that social ties can bring to mental and physical health, such as increased longevity, reduced loneliness, and lower levels of stress~\cite{kawachi2001social, cohen2004social, umberson2010social, thoits2011mechanisms}. Recent studies compared online and offline social ecosystems, studying the formation of online social ties and showing how these ties evolve into social networks~\cite{backstrom2006group, ellison2007benefits, szell2010multirelational, wilson2012review}, including in video games~\cite{zhong2011effects, trepte2012social}. 
Researchers have also been  concerned with understanding how physical and mental factors impact human behavior and performance~\cite{leonard2008richard}. However, little attention has been devoted to study the interplay between social ties and human performance dynamics, which is the subject of this study. 

In the age of big data, humans leave behind traces of their online activity in the form of digital behavioral data, which facilitates our research and bestow us with new data-centric perspectives to study social ties, in addition to established methods like interviews, surveys, or ethnographic observations. In this paper, our main interest is the influence that social ties have on human performance in collaborative team-based settings, more specifically  in Multiplayer Online Battle Arena (MOBA) video games. Understanding the relationship between social ties, in particular preexisting connections within team members, and (individual and/or team) performance is a question of broad relevance across education, psychology, and management sciences~\cite{wuchty2007increasing, guimera2005team, borner2010multi, contractor2013some, Mukherjee2018}, and could lead us to  better understand  what underlies human behaviors in such systems. 
Our research will focus on a popular MOBA collaborative team-based game, \textit{Dota 2}, a rich dataset that will allow us to study millions of players and matches.

Dota 2 is one of the most successful MOBA games: according to the official Dota 2 website, more than ten million unique players participate in the games each month.\footnote{Statistics on Dota 2 Official Website: \url{http://blog.dota2.com/}} 
Dota 2 not only hosts a huge user base but also innately incorporates mechanisms that stress the impact of social ties. Since two opposing teams, each consisting of 5 players, compete against each other,  preexisting friendships are put to test, and strangers are brought together, to collaborate as a team in order to prevail over the rivals. Each player has the autonomy to befriend other players, and these constructed social ties are stored in a \textit{friendship list} on \textit{Steam}, the online game distribution platform that hosts Dota 2 and hundreds of other games and associated communities. In each list, both the time of formation and the actors involved in each dyad are recorded. In this paper, we jointly leverage the behavioral data provided by the log of Dota 2 matches and the social network  data (friendship lists) provided by the Steam community.\footnote{Steam Community Website: \url{https://steamcommunity.com/}}

Motivated by the need for a thorough investigation of the influence of social ties on performance, and in  light of the recent advancement in network science and team science, we analyze our data considering four different perspectives: \textit{(RQ1)} We first look into the actions of individuals that may be affected by the presence of friendship ties in the team. \textit{(RQ2)} Next, we proceed to divide teams into four categories based on both collective experience (experts vs newbies) and performance (high vs low) and analyze them separately. After that, we focus on explaining the dynamics governing performance in short-term sessions of consecutive matches, and propose two additional research questions: \textit{(RQ3)} We plan to understand how individuals perform  in consecutive games when they only play with friends. \textit{(RQ4)} We aim to investigate how social ties affect team performance in consecutive games.

\bigskip

In summary, in this paper we will address the following four research questions (RQs): 

\begin{itemize}
\item[\textbf{RQ1:}] \textit{What is the influence of social ties on individual players' activity?}. We will test whether the presence of social ties affects the activity of individuals within a team. Our hypothesis is that the presence of preexisting friendship ties within a team will increase teammates activity. We will set to test whether there exists a spillover effect by which even individuals who do not have friendship connections with other teammates, but who play in a team where some players are friends among each other, experience such effect. If social ties have an effect, we will also characterize which dimensions of activity it affects.

\item[\textbf{RQ2:}] \textit{What is the influence of social ties on team dynamics?}. We will investigate whether preexisting social ties will affect the performance of the team as a whole. We will further investigate the subsets of teams composed by high/low experience players, and high/low performing players. Our hypothesis is that preexisting social ties improve team performance. We will test whether this is the case, and if so, we will characterize  how  performance  is affected.
\end{itemize}

While the former two questions focus on measuring effects within single matches, the next two questions focus on effects that span over the course of a gaming session (i.e., a nearly-uninterrupted sequence of consecutive matches): 

\begin{itemize}
\item[\textbf{RQ3:}] \textit{What is the influence of social ties on individuals over gaming sessions?}. We will study whether playing game sessions within teams with preexisting social ties affects individuals' short-term activity. We hypothesize that the presence of such ties can mitigate known effects of deterioration in individual performance over the course of the sessions. 

\item[\textbf{RQ4:}] \textit{What is the influence of social ties on teams over gaming sessions?}. We will determine whether short-term  performance of teams as a whole is affected by the presence of social ties. Our hypothesis is again that social ties can influence team performance and mitigate known session-level deterioration effects. 
\end{itemize}

This paper is organized as follows: We will first explain data gathering and preprocessing steps (see Section~\S\ref{sec:data}). In Section~\S\ref{sec:methods}, we will elucidate the methods we employ when answering our four research questions. The results will be presented and discussed in Section~\S\ref{sec:results}. We will also provide an overview of literature concerning social ties, online games, and performance dynamics in Section~\S\ref{sec:related}. In Section~\S\ref{sec:conclusions}, we will conclude our study and shed light on its potential applications  and future extensions. 

% introduce the importance of analyzing social ties and its influence on human performance in the first section. Then we will

%This paper is organized as follows: in Sec. 2 we provide an overview on the data collection methods, data structure, and statistics; in Sec. 3 we explain how we leverage and study social ties in MOBA to study their influence on player performance; in Sec. 4 we investigate the defined research questions and show the results obtained by our analysis; in Sec. 5 we introduce related work in the realm and compare it with our results, and in Sec. 6 we draw the conclusions of our study providing advantages, limitations, and possible future directions.

\section{Data \& Statistics}\label{sec:data}
\subsection{Match log data from Dota 2}

\textit{Defense of the Ancients 2} (Dota 2), is a multi-player online battle arena (MOBA) video game. In this paper, we only study matches consisting of all five human players (as opposed to matches that mix humans and bots, or 1-vs-1 matches). In such matches, two teams each composed by five players compete to destroy the opponent team's fortified home-base known as the ``Ancient''. Each player has the choice to draft a virtual avatar known as a \textit{hero}, to participate in each match. The game is designed with an internal nudging mechanism to foster cooperation, since heroes have complementary abilities (e.g., \textit{Pudge} is popular for its strength, \textit{Sniper} is recognized for agility, \textit{Invoker} is known for intelligence, etc.). Thus, to increase the probability of winning, 
teammates have to coordinate to form balanced teams during the draft phase that precedes each match, and fill various desirable roles (e.g. Carry, Disabler, Support, etc.).  

We acquired match log data of Dota 2 from the OpenDota API.\footnote{OpenDota API:  \url{https://docs.opendota.com/}} This service provides information such as match duration, team members, action statistics of each players, matchmaking type, etc., for millions of Dota 2 matches. The Dota 2 gaming system provides four mechanisms to construct the two opposing teams (matchmaking), i.e., normal match, ranked match, practice 1-vs-1 match, and bot match. Since ranked matches are governed by the Matchmaking Rating (MMR) system, friends cannot freely group together---the goal is to create artificially-balanced opposing teams, thus teams are often composed by random strangers. Practice 1-vs-1 matches and bot matches do not meet our need to evaluate how social ties affect human performance in team-based human environments. Therefore, in our analysis, we only focus on the normal matches, where ten human players participate in one 5-vs-5 match. To win the match, teammates need to collaborate, coordinate, and support each other to harvest resources (e.g., collect gold by killing AI-controlled mobs called creeps), defend their base and towers, attack and defeat the enemies, and destroy their towers and base.

%Although both match types allow various sizes party queues, ranked matchmaking matches have a more strict Matchmaking Rating (MMR) barrier to entry for each player. MMR is an algorithm adapted by gaming system used for calculating each players' skill level. It is based on winning and losing status and will be taken into account in both public and ranked games. In public matches all players' the MMR score will be hidden, whereas in ranked matches MMR score will be explicitly displayed to all members in the match. Only by competing in ranked matches can a player accumulate solo ranked MMR and then appear on the world regional top 200 leader-board. Due to these innate distinctions, we analyzed the two match types separately, however their results were very similar. Thus for simplicity in this paper we consistently conclude only public matchmaking matches' results.

\begin{figure*}[t]
    \centering
    \subfloat[Match Distribution]
{\label{fig:match_distribution}
    \includegraphics[width = .6\columnwidth]{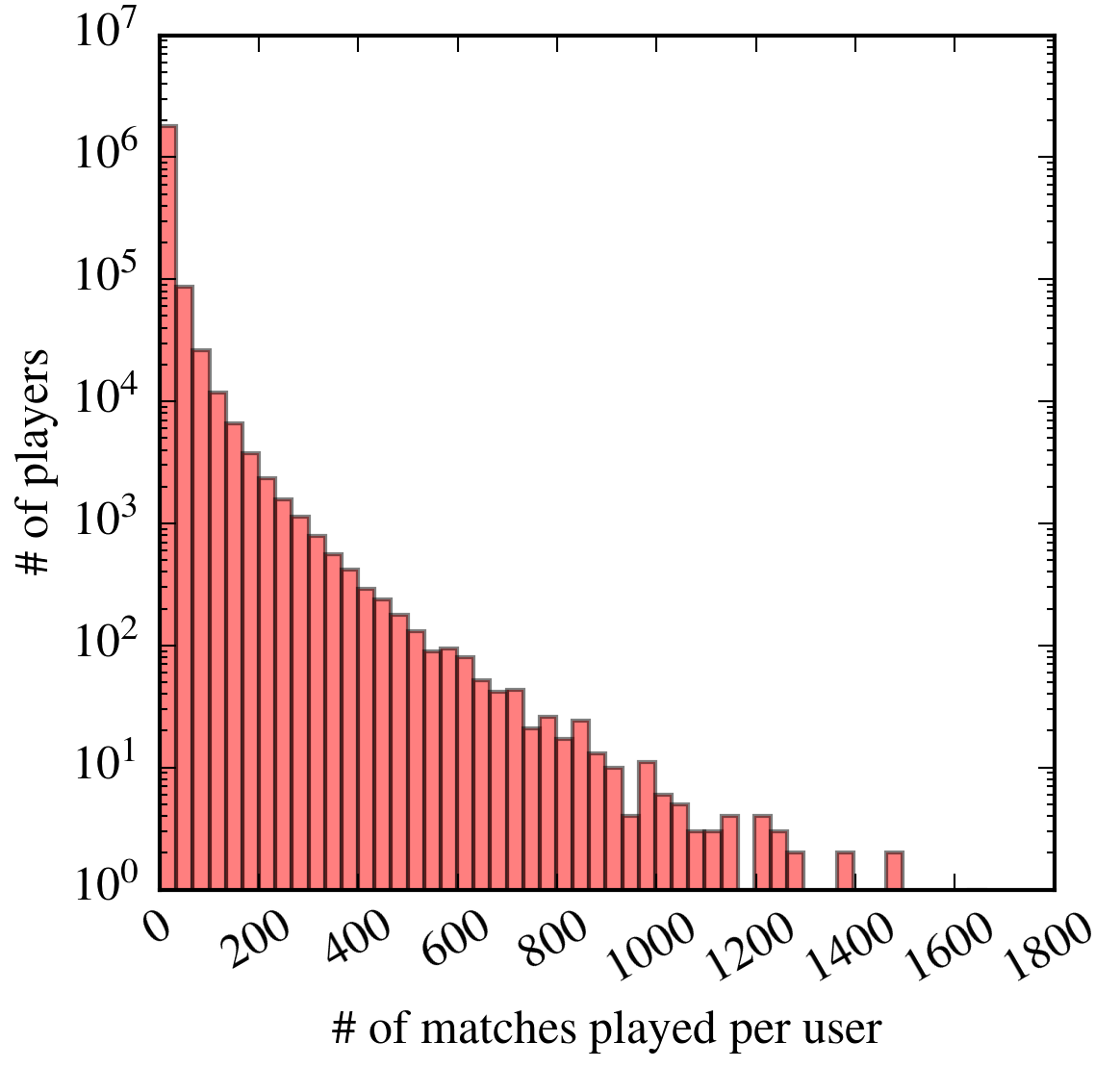}}% 
    \qquad
   \subfloat[Duration Distribution]{\label{fig:duration_distribution}
   \includegraphics[width = .6\columnwidth]{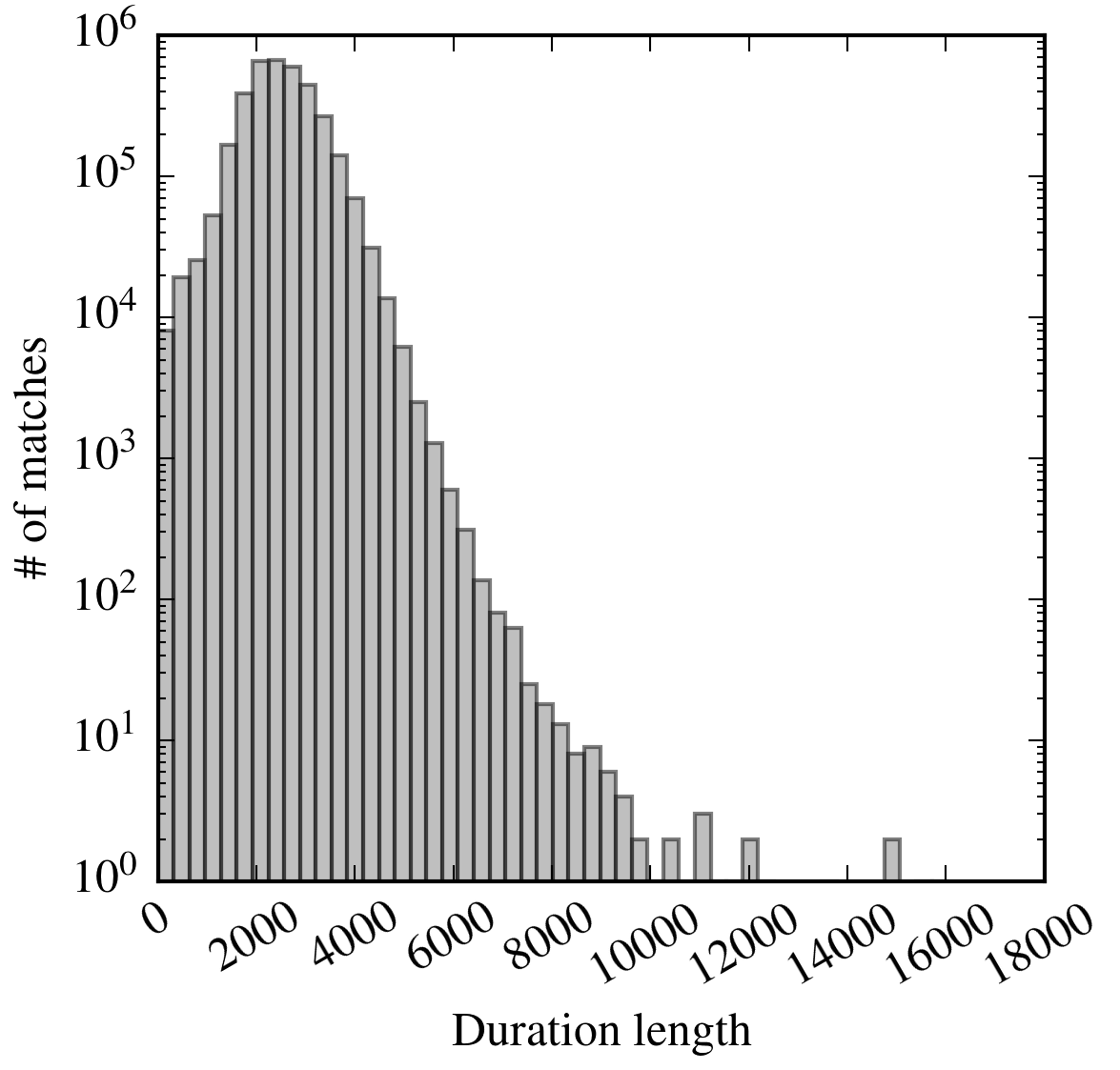}}%
    \qquad
   \subfloat[Individual's Matches Time Gap Distribution]{\label{fig:Matchgap}
   \includegraphics[width = .6\columnwidth]{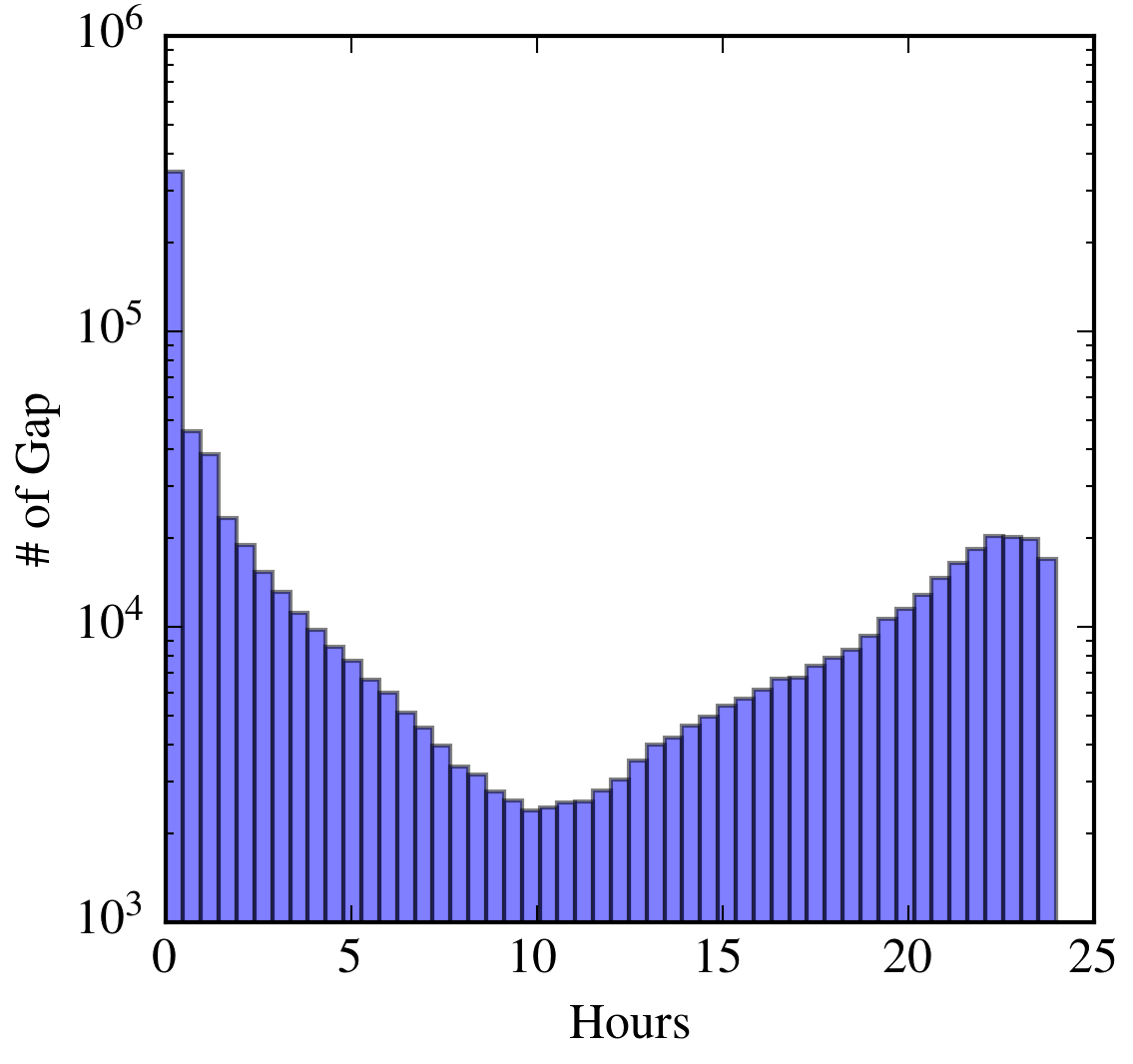}}%
    \caption{\textbf{Distribution plot of our dataset.} (a) Number of matches per player and (b) Duration (in seconds) per match and (c) Duration (in seconds) per match.}%
% \vspace{-0.1cm}
\end{figure*}

Due to data privacy, some users' match records we collected are incomplete. After discarding these unusable match records, our  dataset contains 3,566,804 matches, comprising 1,940,047 unique players, and spans from July 17, 2013 to December 14, 2015. Figure \ref{fig:match_distribution} shows the number of matches per player in our dataset. The average match duration of the matches in our data is about 41.8 minutes, and its distribution is displayed in Figure \ref{fig:duration_distribution}. 

\subsection{Friendship data from Steam}

Steam  is currently the world's largest digital game distribution platform where registered users can not only purchase and manage a variety of games, but also join gaming communities. It is worth noting that the Steam platform and Dota 2 are synchronized via the \textit{convertible Steam account ID} linked to the \textit{Dota 2 player account ID}. Therefore, we are able to connect each Dota 2 player's in-game behavioral data acquired from the Dota 2 API with their friend list (and other account metadata) on Steam. % Thus in our paper, we utilize the connecting IDs to link data between the Steam platform and Dota 2 match logs. 

According to Steam's official Website,\footnote{Steam Official Statistics: \url {https://store.steampowered.com/stats/}} over 10 million players are active on the platform on a daily basis. 
The Steam platform, with its open API, has provided researchers with access to a massive amount of data, that has been leveraged to analyze various aspects of players' behaviors. For instance, \cite{sifa2015large} analyzed play-time related, cross-games behavior of Steam users. \cite{hamari2011framework} proposed analytical abstractions between the different components of game achievements. Other than cross-game behaviors, the gamers' social network provided by the Steam Community has also caught the  attention of the research community: \cite{becker2012analysis} studied the evolution patterns of the Steam community network; \cite{blackburn2012branded} utilized the network structure of Steam to identify cheaters in gaming social networks. Despite of these macro-level analyses, the influence of social ties (i.e., online friendships)  on individual and team performance remains largely unexplored: Therefore, we will utilize the friendship lists of Dota 2 players provided by Steam to reconstruct the player social network and closely examine the impact of social ties on players' in-game performance and behavior. 

The collection of players' friendship lists from the Steam API follows the process of identifying Dota 2 players from match log data. After making sure that each friendship pair was formed before the starting time of each match, we can construct the exact team-wise friendship network structure within each match. We describe teams in each match as a network with 5 nodes, i.e., 5 players are represented by 5 nodes and each pairwise friendship formed before the starting time of the match is recorded as an edge. 

Similar to the Dota 2 API, the Steam API also respects each player's data privacy preferences. We requested friendship information for all 1,940,047 distinct Dota 2 players, and acknowledged that data for 227,045 players was unavailable due to privacy restrictions.

\subsection{Final dataset}

We now finally combine information provided by both the Dota 2 API and the Steam API. To this aim, we make sure that for each team in our final dataset all information about all players' friendships and match actions is openly accessible (i.e., not restricted by privacy settings). Our final dataset therefore contains 954,731 players, 673,864 teams, and 621,629 matches. This is the final dataset used in all our experiments, discussed next. Here, we have records of 365,412 teams consisting of  652,215 unique players who participated in 337,043  normal matchmaking matches. This dataset starts on July 14, 2014 and ends on December 14, 2015. It includes match features, player's individual actions, and social ties related to each team. 
%With the Steam friendship list and complete match log in hand, we embarked on dealing with spontaneous missing data caused by user privacy preference. In the experiment data set we selected teams guaranteeing that 5 member's data are all openly accessible through both Dota 2 API and Steam API.
%while 308,452 teams were in ranked matchmaking games. 

%Their construction methods are introduced in the methods section. 

%\begin{figure*}[t]
%\centering
%\includegraphics[width = 0.95 \columnwidth, clip, trim=1 1 2 1]{./motifs_log2.png}
%\caption{\textbf{34 types of friendship subnetwork structure distribution on public matches and rank matches.}}
%\label{fig:overall_distribution}
%\end{figure*}

\begin{figure*}[t]\centering
\subfloat[Individuals playing with friend(s) \textit{(in-friendship players)} vs \textit{null players} (those players in all-stranger teams).]{\label{fig:overall_in_friendship}\includegraphics[width = .9\columnwidth, clip, trim=4 1 225 1]{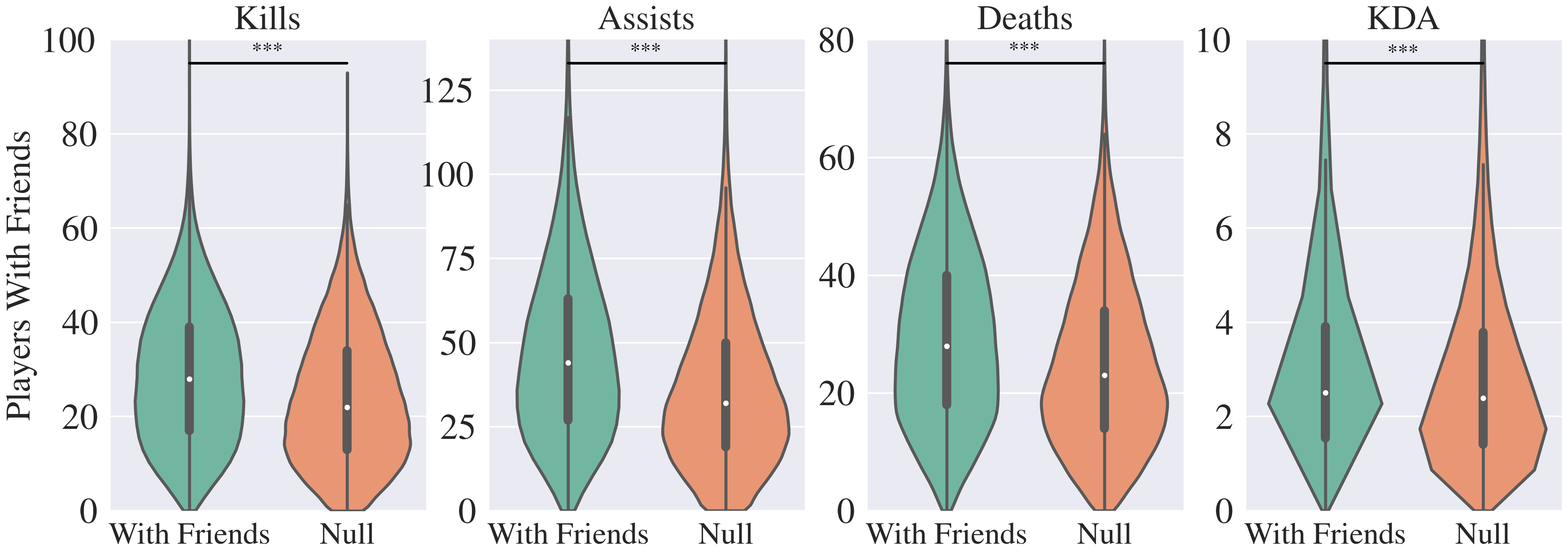}}
\qquad
\subfloat[Individuals playing without friend(s) (\textit{out-friendship players}) vs \textit{null players} (those players in all-stranger teams).]{\label{fig:overall_out_friendship}\includegraphics[width = .9\columnwidth, clip, trim=4 1 225 1]{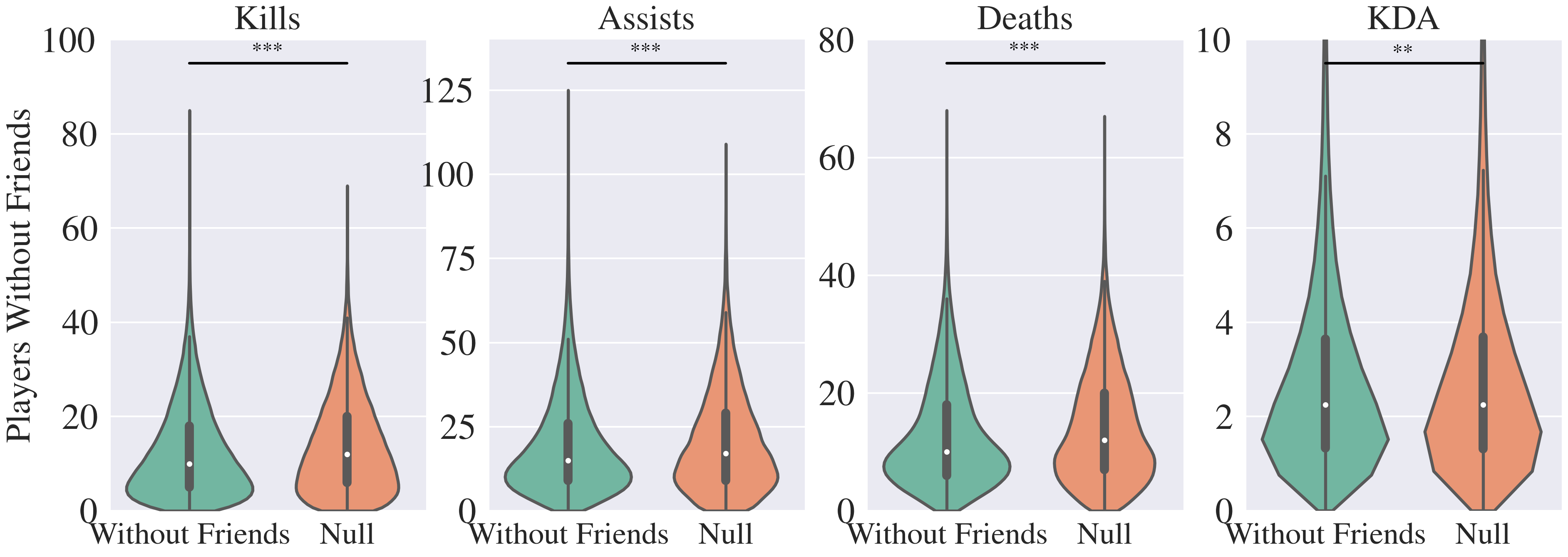}}
\caption{\textbf{Social ties' impact on individuals playing with friends and individuals playing without friends}. Violin plots convey the statistical distribution distinctions between (a) Individuals playing \textit{with friends}, and (b) Individuals playing \textit{without friends}, , in \textit{teams of friends} and \textit{mixed teams}, as well as comparing against \textit{null players} (those who play in all-stranger teams). Shuffled null players are displayed as orange violins (right violin of each pair), observed individuals with and without friends are shown in green violins (left violin of each pair). Stars representing t-test statistical significance are shown in all subplots ($^{***}$ means p-value $<=$ 0.001).}
%\vspace{-.8cm}
\label{fig:Pannoramic_Individual}
\end{figure*}

\section{Methods}\label{sec:methods}

\subsection{RQ1: Overview of the influence of social ties on individuals}

We first introduce the types of social structures we will study. There exist three types of teams in our setting: 

\begin{itemize}
\item \textit{Teams of strangers}, where no preexisting friendship ties exist among any players prior to the match;
\item \textit{Teams of friends}, where each team member has at least another friend in the team; and,
\item \textit{Mixed teams}, where some members have at least on friend among their teammates, while some others do not.
\end{itemize}
  It is worth noting that in our categorization, we do not have a distinct definition for teams that are cliques, i.e., where each player is friend with everyone else, because these instances are exceptionally rare in the data at hand.
Furthermore, we consider individuals playing in teams of friends, as well as in mixed teams, as in conditions  potentially affected by the influence of social ties. Conversely, teams of strangers are used as control groups where players cannot be affected by social ties influence, due to their absence.

% We then hypothesize that friendship imposes no influence on individual players, namely individuals who are under the influence of friendship has identical actions and performance with those who play in teams of strangers. 

Since our first research question focuses on analyzing the influence of social ties on individual players, for each instance of a match, we divide players into three types: 

\begin{itemize}
\item \textit{Null players}, i.e., those playing in a \textit{team of strangers}; due to the absence of social ties, these players are used as null models (thus the name ``null players''), or baselines, to compare and contrast with other player types.  

\item \textit{In-friendship players}, i.e., those playing in \textit{teams of friends}, as well as those playing in \textit{mixed teams}, who have preexisting social ties with some teammates (both sets of players may directly experience social ties' influence).

\item \textit{Out-friendship players}, i.e., those playing in \textit{mixed teams} who do not have preexisting social ties with any of their teammates (yet may indirectly benefit from their teammates' preexisting friendships).
\end{itemize}

In Section~\S\ref{sub:rq1}, we analyze the influence of social ties on players with friends, and players without friends, separately. Take the study on \textit{teams of friends}, for instance: We compare the statistic distributions of performance on in-game actions (kills, assists, and deaths) for \textit{in-friendship players} with that of the \textit{null players}, contrasting the distributions by both statistical analysis and visual analysis (using so-called \textit{violin plots}). We carry out t-test(s) to prove or reject our hypothesis. If the t-test results are statistically significant across all observed distribution pairs, our hypothesis is confirmed and thus we observe an effect of social ties on in-game activity. 
It is worth noting that the data of null players is randomly sampled with reshuffling to yield samples of the exact same size as the samples of in-friendship players. Likewise, we use the same null model strategy to analyze the impact of social ties on out-friendship players.

\subsection{RQ2: Overview of the influence of social ties on teams}

After answering our first research question, we proceed to include not only in-game actions but also performance and experience into our analysis. We use the \textit{kill-death-assist ratio} (KDA) to measure both the performance of individual players and the performance of teams. KDA can be formalized as $(k+a)/max\{1,d\}$, where $k$ is the number of kills, $a$ is the number of assists, and $d$ is the number of deaths of a player (or a team of players) in a given match. 

The teams composed of players with the very high/low experience and very high/low skills are of particular interest for our analysis, provided that they may exhibit noteworthy behavioral patterns: for example, they may exacerbate the effect of social ties' influence in one direction or another. 
To this purpose, we consider the top and bottom 25th percentiles of teams composed by players ranked by average team experience (i.e., number of played matches) and by average team performance (as measured by the average team's KDA score).
To achieve that, we calculate the average KDA and the experience of each team as match-based features. For each match, the average KDA is calculated by averaging all five players’ KDA in the current match (for each of the opposing teams), while the experience is calculated by summing over the number of past matches of all five players until the current match. We select the top 25\% players in each ranking as high-level category and the bottom 25\% as low-level category. Having divided the data in the four categories of teams, in each category we further compare the actions and performance of the whole team, the players with friends (\textit{in-friendship players}) as well as the players without friends in mixed teams (\textit{out-friendship players}) with that of the null players. We compute the difference in each case as $(Y-X)/X$, where $Y$ is the mean of actions or performance of players (or teams) who may be subject to the influence of social ties, and $X$ is the mean of actions or performance of null players  (teams of strangers).
Thus, to summarize, we select four categories of teams as follows:

\begin{itemize}
\item \textit{High Experience \& High KDA}: these are teams composed by players that are in the top 25th percentile by experience (no. played matches) as well as by performance (KDA).
\item \textit{High Experience \& Low KDA}: these are teams composed by players that are in the top 25th percentile by experience (no. played matches) and in the bottom 25th percentile by performance (KDA).
\item \textit{Low Experience \& High KDA}: these are teams composed by players that are in the bottom 25th percentile by experience (no. played matches) and in the top 25th percentile by performance (KDA).
\item \textit{Low Experience \& Low KDA}: these are teams composed by players that  are in the bottom 25th percentile by experience (no. played matches) as well as by performance (KDA).
\end{itemize}

\subsection{RQ3: Influence of social ties on individuals over sessions}

More often than not, individual players tend to complete a sequence of matches rather than a single match before they decide to stop their gaming session. Playing consecutive matches may bring tiredness and boredom to players, which could in turn affect their performance. On the other hand, playing consecutive matches could also help train proficiency. Due to this dichotomy, this aspect warrants further investigation.
Therefore, we formalize consecutive playing patterns as individual gaming sessions. Provided that we don't know exactly when players start or interrupt a gaming session, we need to infer such sessions from the start/end times recorded in each match metadata.

We set 1 hour as the threshold to split gaming sessions:  if the time gap between the end of a match and the beginning of the next match, for each player, is shorter than one hour, we assume that these two matches belong to the same gaming session; otherwise, two separate gaming sessions are extracted. We calculate all the time gaps---the time intervals between the end of a match and the beginning of the subsequent match---for each player and concatenate all players' time gaps together. Fig. \ref{fig:Matchgap} shows the distribution of time gaps that is less than 24 hours in our dataset: amongst these 84K time gaps, the median is 1.265 hours, supporting our choice of 1 hour threshold to split sessions. Each gaming session consists of a list of consecutive matches ordered by their starting time. Such sequence index in a  session is named as \textit{match position}. For example, in a  session of four matches, the first match is called \textit{match in position one}, and the last match is referred to as \textit{match in position four}.

To isolate the effect of social ties on gaming sessions, in RQ3, we focus on individuals who only play with friends throughout the entire  session. This will allow us to reduce the variability that may arise in case of inclusion of mixed sessions where users played  games both with and without friends. Of course, this filter also reduces the number of  sessions  suitable for analysis.
We use two strategies to analyze social ties’ impact on these individuals: 

\begin{itemize}
\item 
First, we study the individuals' KDA trajectories throughout the gaming sessions. We only utilize gaming sessions with length 1 to 4, as data about sessions of length larger than 4 is very sparse (for reference, a gaming session of length 4 usually spans between 3 and 5 consecutive hours of uninterrupted playing; anecdotally, in our data, we observe isolated instances of sessions that last  up to 20 consecutive hours). 
For gaming sessions with different length, we separately aggregate the KDA on each match position, and use separate line plots to visualize the trajectories over the course of the sessions. Then, we randomly reshuffle the sequence of matches in all gaming sessions, and reconstruct the trajectories based on the shuffled data---this is used as a randomized null model. By  comparing the trajectories in  sessions with original match positions against trajectories in sessions with randomized match positions, we exclude the possibility that any emerging trend  is a produced just by chance. 

\item
Second, we compute the KDA difference between the last and the first match of a session, expressed as $(Y-X)/X$, where $Y$ is the KDA performance of the last match in a gaming session and $X$ is the KDA performance of the first match in a gaming session. This measure is adopted to capture the variation of overall performance throughout the whole session, i.e., the overall size of such an effect.
\end{itemize}

\subsection{RQ4: Influence of social ties on teams over sessions}
\label{sub:rq4def}
In the previous section, we introduced the notion of gaming sessions of individual players. Here, we define a team’s gaming session as the average gaming session of its 5 individual players. For example, for a given team in a given match, three players may be playing the first match of their session, two players may be playing the second match of their session, and one player may be playing the fourth match of their session: in this case, the average session length for this team would be $(1+1+1+2+4)/5=1.8$.
Therefore, due to the employed averaging strategy, the length of a team’s gaming session will be expressed in a ranges, i.e., gaming sessions of length [1-2), [2-3) and [3-4). Sessions of average length greater than 4 are exceptionally rare and therefore excluded from our analysis.

To answer RQ4, we analyzed the kills, assists, and deaths as well as KDA performance of players subject to the influence of social ties, considering team’s gaming sessions. We use one pair of violin plots to visualize each type of action or KDA’s distributions of \textit{in-friendship players} versus \textit{out-friendship players} in each team. We then use t-test(s) to verify the statistical difference of each pair. To investigate the trend of teams’ actions and performance when the length of gaming session increases, we organize the plots by comparing each type of action (or KDA) across teams of different session lengths. Consider, for example, the plot tracking team kills (take a glance at Fig.~\ref{fig:team_session_kills}): from left to right, the first pair of violin plots belongs to teams with avg. session length 1-2, the second pair of violin plots belongs to teams with avg. session length 2-3, and the third pair of violin plots belongs to teams with avg. session length 3-4.

%%%%%%%%%%%%%%%% Results %%%%%%%%%%%%%%%%%%

\begin{figure}[t]
	\centering
	\includegraphics[width = .95\columnwidth]{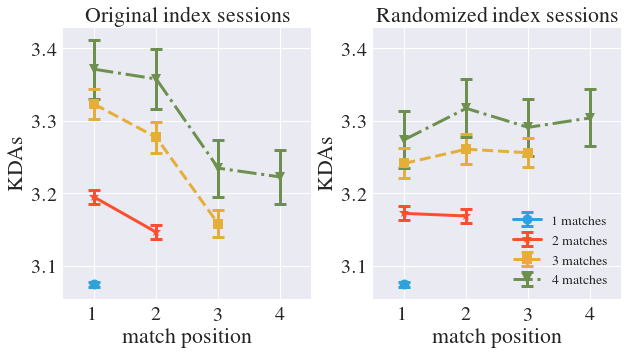}
	\caption{\textbf{KDA trajectories of individuals who only play with friends throughout the entire gaming session.} The left plot shows the actual data suggesting the presence of individual performance deterioration over the course of gaming sessions. The right plot shows the reshuffled null model where the effect of match position is disrupted (therefore, the lines are expected to become flat).}
% 	\vspace{-.3cm}
    \label{fig:individual_session}
\end{figure}

\begin{figure}[t] 
\centering
\includegraphics[width=0.5\textwidth]{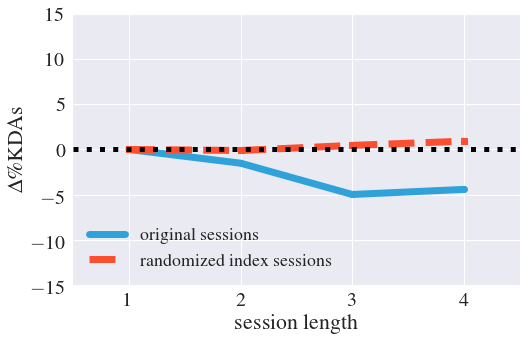}
\caption{\textbf{Changing rate of KDA performance in sessions of different lengths by individuals who only play with friends throughout the entire gaming session.} This plot shows the KDA change percentage of last game in the session from the first game in the session from in-friendship players (those who played the entire session with some friend(s) in their team).}
% \vspace{-.3cm}
\label{fig:individual_session_percentage}
\end{figure}

\section{Results}\label{sec:results}

In this section, we present the results of our analysis aimed to address the research questions defined above. We present our summary in four parts, corresponding to the four proposed research questions. For RQ1 (\S\ref{sub:rq1}), we provide an overview of the interplay between individual players' activity and the effect of social ties. To answer RQ2 (\S\ref{sub:rq2}), we focus on friendships' influence on team's actions and performance by categorizing teams according to average experience and KDA performance. We explore RQ3 (\S\ref{sub:rq3}) by  identifying performance dynamics of individual players who only play with (and without) friends throughout their entire gaming sessions. Lastly, for RQ4 (\S\ref{sub:rq4}) we extend our analysis to actions and performance trajectories of teams within gaming sessions. For each research question, we will aim at validating or rejecting the hypotheses formalized above.

\begin{table*}[t]
\centering \small
\begin{tabular}{@{}c|l|lll|l@{}}
\toprule
\multicolumn{1}{l|}{Team Category} & Condition & Kills & Deaths & Assists & KDA \\ \midrule\midrule
\multirow{3}{*}{\begin{tabular}[c]{@{}c@{}}Low Experience\\  Low KDA\end{tabular}} & Whole Team & 87\% & 129\% & 19\% & 79\% \\
 & In-Friendship & 96\% & 15\% & 27\% & 474\% \\
 & Out-Friendship & 79\% & 97\% & 9\% & 990\% \\
 \midrule
\multirow{3}{*}{\begin{tabular}[c]{@{}c@{}}High Experience\\  Low KDA\end{tabular}} & Whole Team & 109\% & 125\% & 18\% & 89\% \\
 & In-Friendship & 185\% & 209\% & 54\% & 437\% \\
 & Out-Friendship & 48\% & 56\% & -1\% & 392\% \\
 \midrule
\multirow{3}{*}{\begin{tabular}[c]{@{}c@{}}Low Experience\\  High KDA\end{tabular}} & Whole Team & 17\% & 23\% & 94\% & -36\% \\
 & In-Friendship & 25\% & 35\% & 109\% & -32\% \\
 & Out-Friendship & 1\% & 8\% & 79\% & -28\% \\
 \midrule
\multirow{3}{*}{\begin{tabular}[c]{@{}c@{}}High Experience \\ High KDA\end{tabular}} & Whole Team & 39\% & 52\% & 151\% & -32\% \\
 & In-Friendship & 67\% & 85\% & 208\% & -26\% \\
 & Out-Friendship & 9\% & 13\% & 89\% & -23\% \\ \bottomrule
\end{tabular}
\caption{Percentage difference of 4 categories of teams' action/performance compared with null model of all-stranger teams.}
\label{tab:team_percentage}
\end{table*}

\subsection{RQ1:  Influence of social ties on individual players' activity}\label{sub:rq1}

%There are 652215 players appeared in the our public matchmaking matches dataset. 
To study how social ties influence players' activity in the game, we compare \textit{teams of friends} and \textit{mixed teams} against \textit{teams of strangers}. We call \textit{teams of friends} and \textit{mixed teams} as teams with preexisting social ties. \textit{Teams of strangers} are consider as our null model since no social ties exist in this condition. In other words, we consider players in teams of strangers as our control group, while players in teams with preexisting ties are assembled as observed conditions in the following analysis aimed to tackle the first research question. 

We evaluate observed players' actions (kills, assists, and deaths) against the null model, i.e., comparing individual's action patterns under the presence of social ties in contrast with all-strangers team scenario. Data of players in all-stranger teams are randomly shuffled and then under-sampled (or over-sampled) to match the number of players in our observed conditions data. 

As long as there exists some social tie in the team, individual players could be divided into two types, namely individuals playing with some friend(s) and individuals playing without any friend(s). We named the first group \textit{in-friendship players}, and the second group as \textit{out-friendship players}. Our hypothesis is that if social ties have some form of influence on players' activity, \textit{in-friendship players} will experience this effect directly (since these are the players who are playing with some friends), while \textit{out-friendship players} may experience it indirectly, even without playing with friends, yet by playing with teammates who are friends among each other.

The plot for players with friends (a.k.a., in-friendship) is shown in Figure \ref{fig:overall_in_friendship}. The plot for players without friends (a.k.a., out-friendship) can be found in Figure \ref{fig:overall_out_friendship}. Note that in Figure \ref{fig:overall_in_friendship}, we use ``With Friends'' to label the distributions associated with in-friendship players, while in \ref{fig:overall_out_friendship}, we use ``Without Friends'' to mark the distribution associated with out-friendship players. Stars in the all plots represent t-test statistical significance obtained by comparing observed conditions versus the null model (\textit{null players}, i.e., those in all-stranger teams where no social ties exist).

By inspecting Figure \ref{fig:overall_in_friendship}, we observe that, in comparison to null players (players in all-strangers teams), individuals playing with friends have higher number of kills and assists. However, deaths also arise along with kills and assists. In other words, they are more engaged and active in the game, which leads to an increased number of in-game actions, both positive (kills and deaths) and negative (deaths). This may suggest that players with friends in their team may tend to adopt more aggressive or impulsive strategies. 
%The KDA performance of players with friends are more centralized to the median of 2.3. The median of kills and assists and deaths of n-friendship players are 7.25, 11.5 and 7.5. 
Figure \ref{fig:overall_out_friendship} shows a reverse patterns: contrary to players with friends (in-friendship players), players without friends (out-friendship ones) have relatively fewer actions in comparison with null players. Such decrease suggests that players in the out-friendship condition may tend to act in the best interest of themselves, adopting a more conservative play style. Alternatively, they may also experience being left out of the coordination and therefore being exposed to less game action, thus having fewer opportunities throughout a match to accomplish both positive and negative actions. 

% By doing so their KDA performance lingers around the median of above 2.15. Their median for kills and assists and deaths are 6.5, 10.5 and 7. By comparing figure \ref{fig:overall_in_friendship} with figure \ref{fig:overall_out_friendship}, we observe that given friendship existing scenarios players with friends are obviously motivated in actions and elevated in performance, whereas players without friends are more analogous to the null players but may still be slightly promoted. Their statistics show that although for players with friends more deaths occurred, but the sharper boost on kills and assists mitigates the loss of death, resulting in better KDA performance than those without friends. 

\subsection{RQ2:  Influence of social ties on team dynamics}\label{sub:rq2}

To better understand how social ties impact teams as a whole, we divide teams into four categories: \textit{(i) High Experience \& High KDA}, \textit{(ii) High Experience \& Low KDA}, \textit{(iii) Low Experience \& High KDA}, and \textit{(iv) Low Experience \& Low KDA} teams. Moreover, by comparing them with our null model, all-stranger teams, we analyze the actions (kills, deaths, assists) and performance (KDA) of each category. To compare the two groups we calculate the percentage-difference of actions as: $(Y-X)/X$, where $Y$ is the mean of actions/performance of observed players in teams with social ties, and $X$ is the mean of actions/performance of the null model. In Table \ref{tab:team_percentage}, we report the results we obtained.

We can observe that low-experience \& low-KDA teams have kills, assists, and deaths actions all higher than those of all-strangers teams. In terms of having positive percentage gain of actions, in-friendship players are the biggest winners since they almost double the amount of kills and also have a 27 percent raise on assists. However, out-friendship players turn out to be the largest beneficiary of KDA performance. Although in-friendship players gained a 4.74 times performance boost by collaborating with friends, the out-friendship players received twice as much benefits by an indirect effect. Such effect is not consistent with regard to high-experience \& low-KDA teams, whose in-friendship players are the biggest gainer in all actions as well as KDA performance. Moreover, their out-friendship players show increased unwillingness to help other teammates since their assists are even less than null players. For the remaining two categories with high KDA, in-friendship players have the largest actions percentage gain, but the whole team is the biggest loser in terms of KDA performance. 

When comparing across the four categories, players with friends, and the whole team, have consistently positive gains on kills, deaths, and assists. We observe that in-friendship players in high-experience categories double the actions in comparison to their low-experience counterparts. However, it is not the case for out-friendship players, namely their actions are not obviously affected by experience difference. This reveals that, for players with friends, experience boosts activity but it is not effective on performance. We also observe that for high-KDA teams, their KDA drops drastically when exposed to a team with preexisting social ties. The sharpest rate of decline goes to the whole teams' performance, followed by the decline rate of 20 percent concerning the in-friendship players and out-friendship players. While for low performing teams, the KDA improves drastically when compared with teams of all strangers, such that on a whole team level their performance rose by almost 80 percent, while in-friendship players exhibit over 4 times higher KDA. %For starting teams with bottom experience and bottom KDA, the performance boost of out friendship players trump all other players. 

As a summary, low-performance players, regardless of their experience, exhibit the highest gains in KDA when preexisting social ties are present in the team. Conversely, high-skill players exhibit significant decreases in KDA, regardless of their experience, when social ties are present in the team. In other words, playing with friends benefit almost exclusively low-performance players, who drag down the performance of their better-skilled friends.

\subsection{RQ3: Influence of social ties on individuals over sessions}\label{sub:rq3}
To answer RQ3, we focus on analyzing the impact of social ties on individuals who only play with friends throughout the entire gaming session. %These players are selected based on whether they remain as an in-friendship player throughout the entire gaming session. 

Figure \ref{fig:individual_session} displays the KDA performance trajectory of individuals who only play with friends throughout the entire gaming session: The left plot shows the actual data suggesting the presence of individual performance deterioration over the course of gaming sessions. For example, for sessions of length 3, the average KDA in the first match of such sessions is above 3.3, while the average drops to below 3.2 in the third and last match of such sessions. This effect is visible across the three conditions of session length greater than one.
Then we verified our findings via randomization to exclude the possibility that the performance deterioration phenomenon was created by chance (random effect). The right plot of Figure \ref{fig:individual_session} shows the reshuffled data, where the effect of match position is disrupted: as we would expect, the lines flatten out suggesting that indeed the position of a match in a session has an effect on performance, corroborating the performance deterioration hypothesis in line with recent research results~\cite{sapienza2017performance, sapienza2018non, sapienza2018individual}.

To quantify the effect size of such performance deterioration, we computed the percentage-change in KDA between  the last and the first game in each session. Figure \ref{fig:individual_session_percentage} shows the percentage change in KDA performance for sessions of length 1, 2, 3 and 4. The greater effect size is experienced for sessions of length 3 and 4, where the drop in KDA is approximately 5\% (the randomized model as expected shows a flat line suggesting the absence of such effect in the null model).
Summarizing: Results in Figure \ref{fig:individual_session} and \ref{fig:individual_session_percentage} reveal that players who only play with friends in a gaming session display an apparent trend of performance deterioration. However, such strong trend was not apparent for players who only play without friends and players who only play with strangers in a gaming session.

\begin{figure*}[t]\centering
\subfloat[Kills of In-friendship players vs Out-friendship players ]{\label{fig:team_session_kills}
\includegraphics[width = .95\columnwidth]{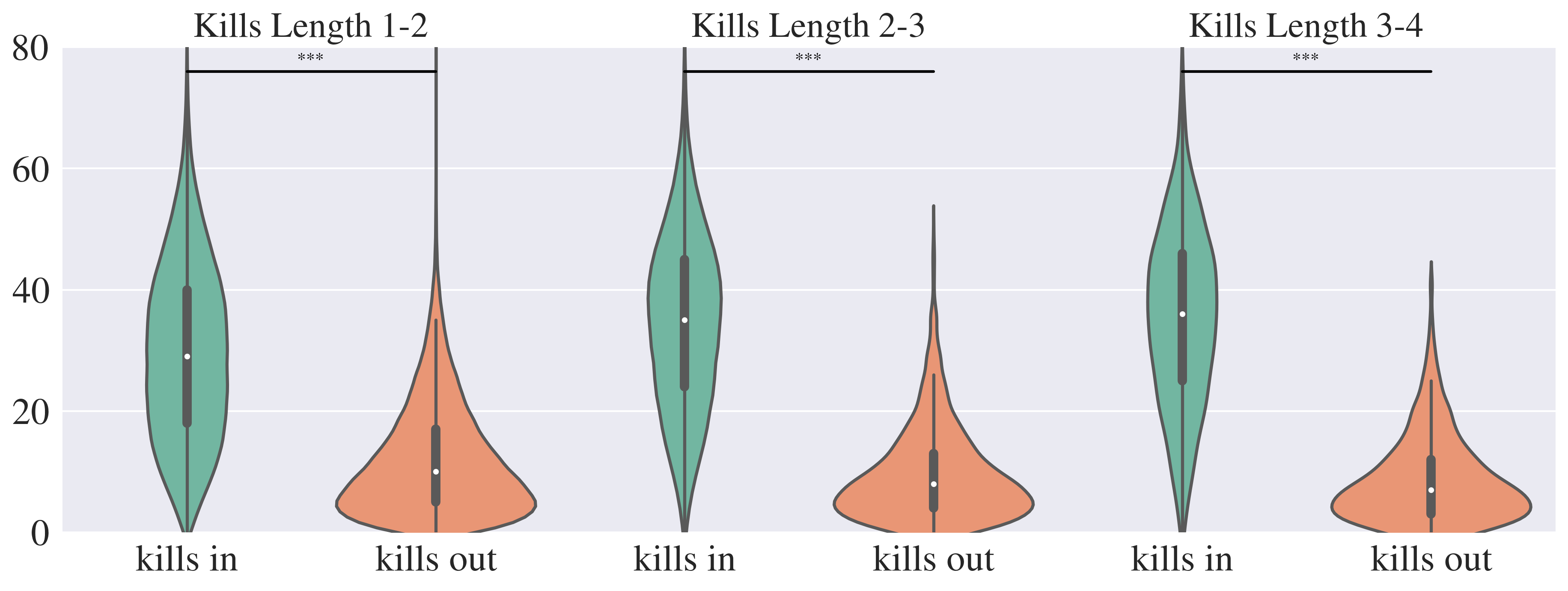}}
\qquad
\subfloat[Assists of In-friendship players vs Out-friendship players ]{\label{fig:team_session_assists}
\includegraphics[width = .95\columnwidth]{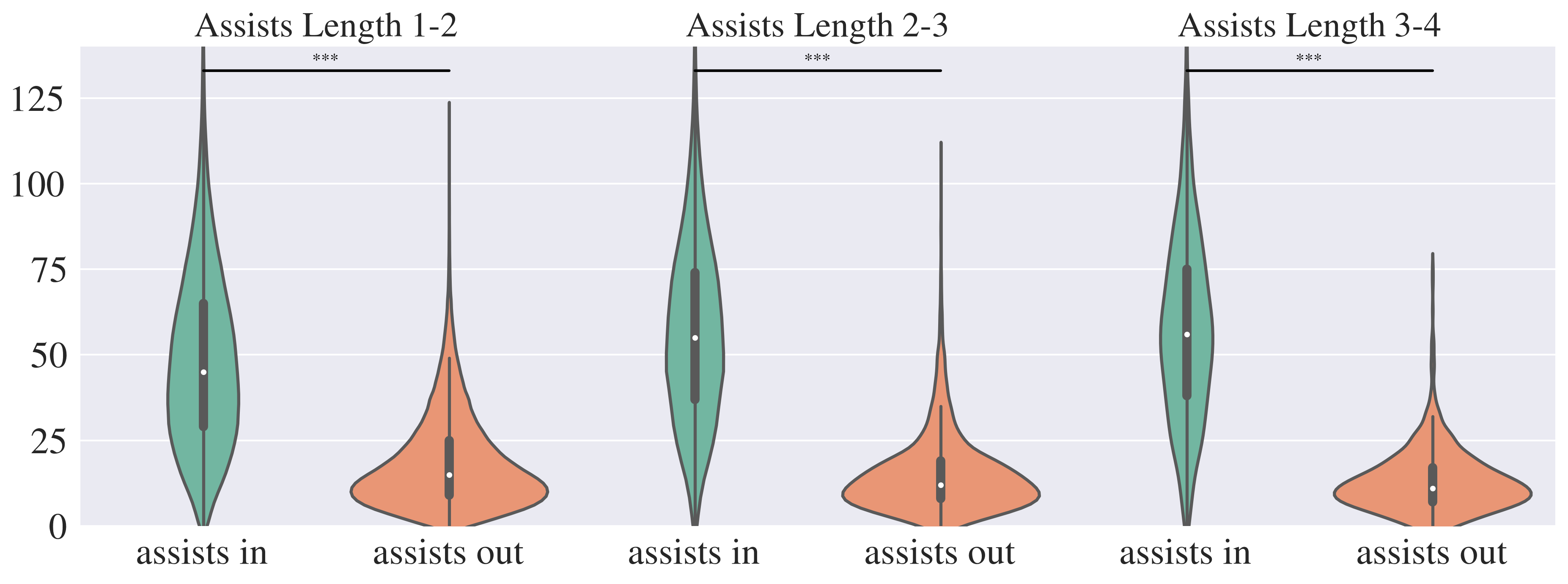}}
\qquad
\subfloat[Deaths of In-friendship players vs Out-friendship players ]{\label{fig:team_session_deaths}
\includegraphics[width = .95\columnwidth]{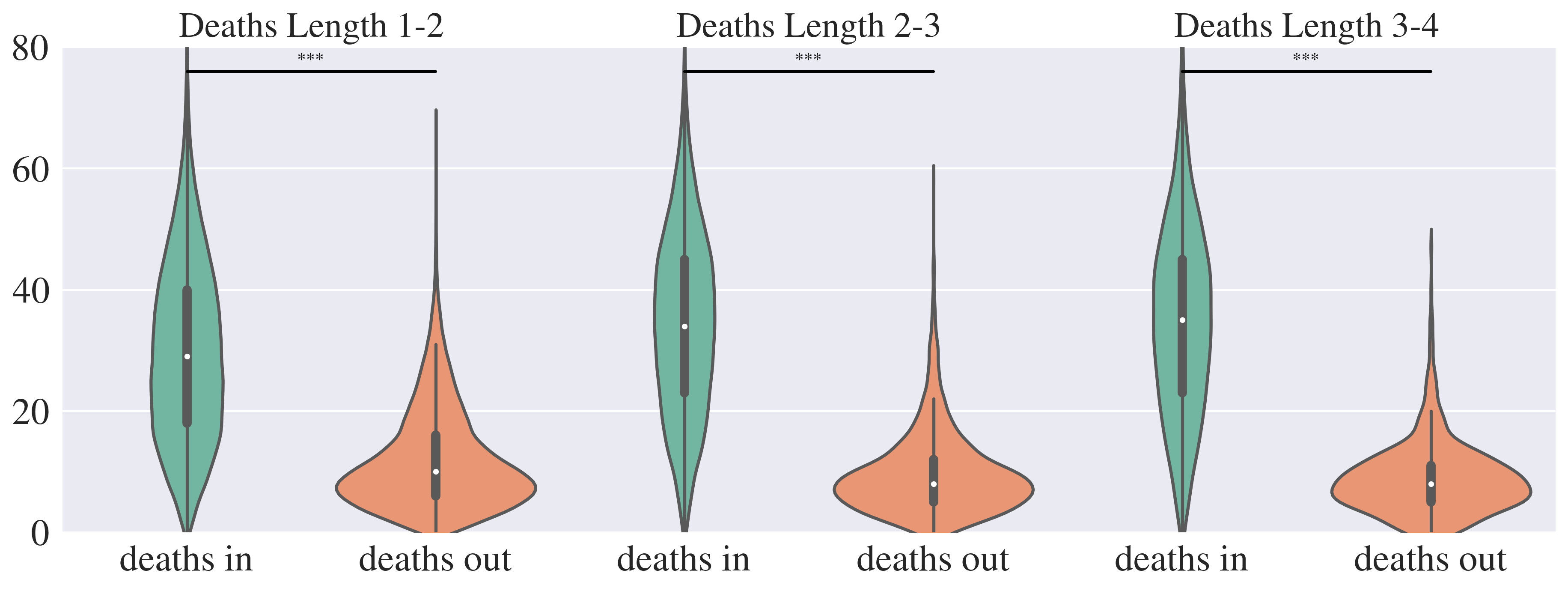}}
\qquad
\subfloat[Average KDA of In-friendship players vs Out-friendship players ]{\label{fig:team_session_KDA}
\includegraphics[width = .95\columnwidth]{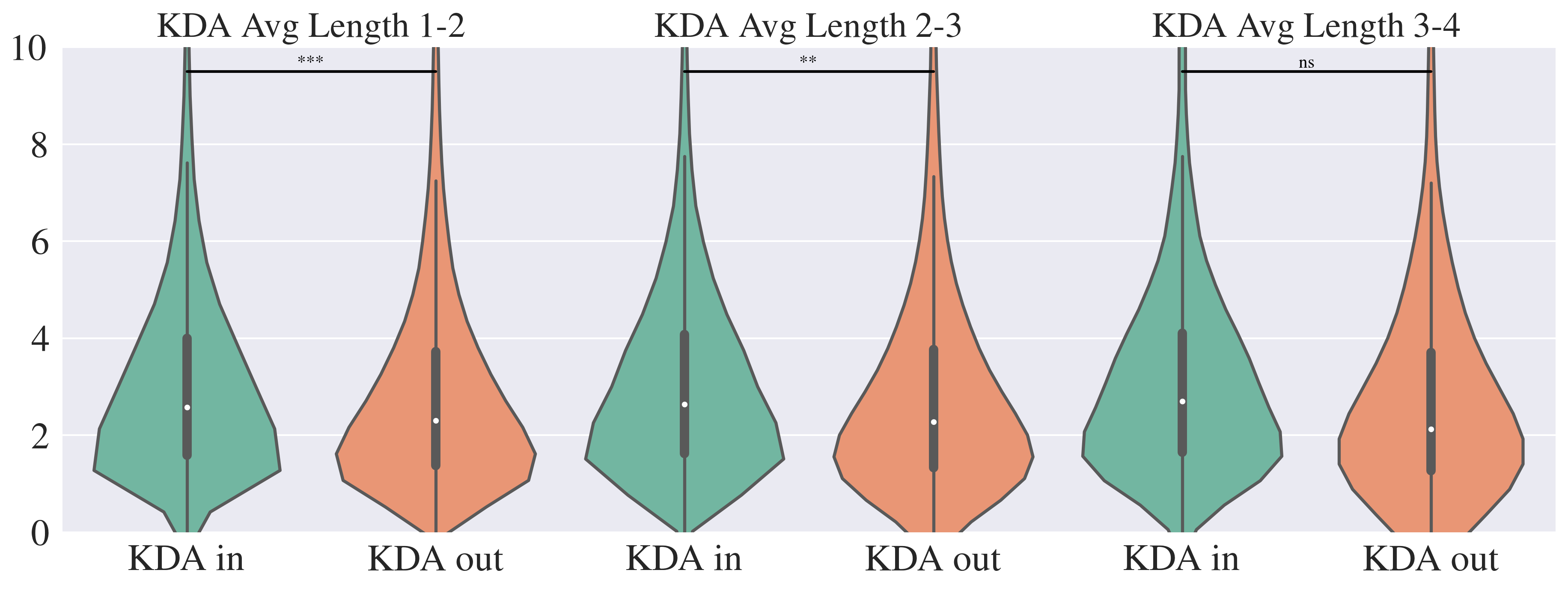}}
\caption{\textbf{Social ties' impact on teams over gaming sessions.} Violin plots convey the statistic distributions between players with friends and players without friends in teams with preexisting social ties. Four aspects were examined: (a) kills, (b) assists, (c) deaths, and (d) KDA performance. In-friendship players' data are displayed as green violins (left violins), while out-friendship players' data are shown in orange violins (right violins). Stars represent t-test statistical significance ($^*$ means p-value$<$0.05, $^{**}$ means p-value$<$0.01, $^{***}$ means p-value$<$0.001, $ns$ means not-significant).}
% \vspace{-.75cm}
\label{fig:Team_Session}
\end{figure*}

\subsection{RQ4: Influence of social ties on teams over sessions}\label{sub:rq4}
%Having grasped the gist of behavioral patterns of both individuals and teams, 
We now analyze how actions and performance change in relation to different session positions in presence of social ties. Here we focus on the analysis of actions and performance over the entire team.%We first begin with overall team's actions and performance in RQ3 and then proceed to individual level in RQ4. 

As mentioned in Section~\S\ref{sub:rq4def}, the average gaming session's length of a team falls into three ranges, namely [1-2), [2-3), and [3-4). Data of sessions beyond that average length are excluded due to high sparsity and low significance.
Furthermore, we concentrate exclusively on teams with social ties. In Figure \ref{fig:Team_Session}, we compare individuals playing with friends (in-friendship players) with individuals playing without friends (out-friendship players). Figure \ref{fig:team_session_kills}, Figure \ref{fig:team_session_assists}, Figure \ref{fig:team_session_deaths}, and Figure \ref{fig:team_session_KDA} each convey the kills, assists, deaths, and average KDA performance distributions of in-friendship players vs out-friendship players, on different session positions. All the distributions of out-friendship players are labeled as ``out'', whereas in-friendship players are labeled as ``in''.

Our results overall reveal that, as time goes by throughout a gaming session, individuals playing with friends have gradually increasing kills, assists, deaths and KDA scores, while individuals playing without friends have decreasing actions and performance. Such results suggest that, except for experience, the presence of social ties in a team can also help players mitigate performance deterioration over the short term throughout a gaming session.

%%%%%%%%%% Insert Related work here %%%%%%%%%%%%%%
\section{Related Work}\label{sec:related}

\subsection{Social ties and performance}

% Human behaviors are not always driven by economic rationality. Sometimes supposedly irrelevant factors like social ties may nudge you into deviating behavior patterns and lead to different performance results~\cite{leonard2008richard,thaler2015misbehaving}. 
% Social ties of individuals are multidimensional including large variety of information-carrying connections between people
% In this paper, we focus on analyzing the connections on ``friendship lists'' formed on the Steam platform. Although these connections may incorporate romantic relations, blood relationship or leader-member relationship, but to be consistent with the name given by Steam, we call the connections in Steam's ``friendship lists'' as friendships in this paper.

Our research is solely based on objective measurements of behavioral data from Dota 2 players where connections between players are studied as social ties. However, most existing social tie studies heavily rely on interviews, surveys, or ethnographic observations to collect self-reported relational data from study participants where each individuals are asked about their proximity to and friendship with others.
\cite{eagle2009inferring} collected both self-reported data and mobile phones usage behavioral data from 94 subjects. The authors revealed that self-reports are biased toward recent and more vivid events. 

Some empirical studies examined the effect of friends (vs. strangers) on individual and team performance. 
% 
%Despite the massive amount of research studying the role of social ties in communities (groups, teams, etc.) and society at large, little attention has been devoted to study their interplay with other human behavioral dynamics. Of particular interest is the influence that social ties have on human performance in  collaborative team-based settings.
% 
In the context of business psychology and organizational development management, for example, workplace ties have for long been the subject of debate.
~\cite{chung2018friends} proposed that relative to acquaintance groups, when friends work together, coordination will be improved through increased collaboration, communication, and conflict management; motivation will be increased via increased commitment, goal-setting, and goal pursuit. Their results revealed that friendship has a significant positive effect on group task performance.  Although many scholars and practitioners have assumed that friendships lead to desirable organizational outcomes,~\cite{pillemer2018friends} explained the downsides associated with workplace friendships.

Researchers have also investigated the influence of social ties, especially peer-effects in academic environments, where youngsters in classrooms become ideal study targets. ~\cite{wentzel1997friendships}  reached a conclusion that aspects of peer relationships are related to GPA indirectly, by way of significant relations with pro-social behavior i.e., the behavior of helping other children learn. Since their study,~\cite{zimmerman2003peer} found that students in the middle of the SAT distribution may have somewhat worse grades if they share a room with a student who is in the bottom 15 percent of the verbal SAT distribution. \cite{ding2007peers} investigated that high-ability students benefit more from having higher-achieving schoolmates and from having less variation in peer quality than students of lower ability. \cite{burke2013classroom} found that low ability students benefit about twice as much from an increase in the share of top-quality peers as they do from an increase in the share of low ability peers. Middle students will benefit from mixing with top-quality peers as well.

%Social ties have also been studied in individual and group behavior in sport teams.~\cite{hellandsig1998motivational} studied the motivation of high performance teenage athletes who performed in explosive, endurance, and team sports. They found that high performance in endurance sports was best predicted from high scores on goal orientation and the importance of friendship attributed to self. ~\cite{lusher2010application} used social network analysis methods, and found that ability is not related to nomination as a friend, and that the best players are friends with one another, therefore cluster together in one part of the network.
Our paper focuses on analyzing the effect of social ties on the performance of online games. ~\cite{leenders2016once} pointed out that while much progress has been made in detailing different types of team processes, empirical evidence of their predictive validity is generally underwhelming and they pointed to the need for a more specific temporally rich theoretical formulation of process. While identifying these critical rarely analyzed aspects, in this paper we used gaming sessions to capture behavior patterns considering temporal limitations.

\subsection{Social ties in games and virtual teams}
Teams science is essential to organizations, informal groups and individuals \cite{wuchty2007increasing, guimera2005team, borner2010multi, contractor2013some, Mukherjee2018}. Considerable attention has been paid to teams across a range of interdisciplinary  challenges. However, the factors affecting team performance in complex, realistic task environments remain yet scarcely understood, both in theory and in practice. In this paper, we focus on individual and team engagement in online games, specifically in a MOBA game called Dota 2. 

% Research has been conducted on World of Warcraft (WoW), which is one of the most popular Massively Multiplayer Online Role Playing Games (MMORPG). ~\cite{ducheneaut2006alone} described a phenomenon in WoW where players tend to be “alone together”, meaning players are surrounded by others but not necessarily actively interacting with them. They also assess the players’ social experiences in longer-lived player associations (the guilds). Their results suggested guilds can encourage players to play more often and more regularly in short term but suffers from long-term maintenance. 
% ~\cite{martonvcik2016world} found in the sample consisting of 161 WoW players, that players experience a significantly lower degree of loneliness and social anxiety in online than in real world. The lower degree of loneliness experienced was also associated with playing with friends and known people, with guild membership, as well as frequent communication with teammates through VoIP (Voice over Internet Protocol) services. Their results suggest that WoW is a highly social environment that encourages cooperation, communication and friendship.
Research has been conducted on a Massively Multiplayer Online Role Playing Games (MMORPG), Dragon Nest~\cite{wax2017self}, which revealed that self-assembled teams form via three mechanisms: homophily, familiarity, and proximity; the authors show that successful and unsuccessful teams were homogeneous in terms of different characteristics, but successful teams are more often formed based on friendship than those unsuccessful teams. 

A recent work by Mukherjee \textit{et al.}~\cite{Mukherjee2018} further corroborated this hypothesis by analyzing both sports (football, cricket, baseball) and esports (Dota 2) suggesting that success shared in prior team experiences is an excellent predictor of future team success. This would imply that social ties and prior experience between team members play a more important role in successful team dynamics than the so-called ``superstar effect''---the idea that well-constructed teams can be outperformed by poorly-constructed teams that have a superstar player whose skills are far better than everyone else.

Other research investigated \textit{social-network games}. A social-network game is a type of online game that is played through social networks. They typically feature multiplayer gameplay mechanics. Social-network games were originally implemented as browser games. As mobile gaming took off, the games moved to mobile as well. 
% ~\cite{domahidi2014dwell} found that social online gamers do not differ significantly from other gamers or non-gamers regarding the number of their good friends. 
However, we found a significant impact of social online gaming frequency on the probability of meeting exclusively online friends. Different social motives played an important role for modality switching processes. Players with a pronounced motive to gain social capital and to play in a team had the highest probability to transform their social relations from online to offline context. We found that social online gamers are well integrated that they use the game to spend time with old friends as well as to recruit new ones.

\subsection{Performance deterioration dynamics}
A recent research thread is concerned with quantifying the temporal dynamics of performance in techno-social systems.
Short-term deterioration of individual performance was previously observed in real world (offline) tasks. Recent studies investigate this phenomenon by drawing a parallel with online platforms: research shows that the quality of comments posted by users on Reddit~\cite{singer2016evidence}, the answers provided on StackExchange question-answering forums~\cite{ferrara2017dynamics}, and the messages written on Twitter~\cite{kooti2016} and Facebook~\cite{kooti2017understanding} decline over the course of an activity session. Other than individual online behaviors, short-term deterioration effects has also been found in virtual teams in MOBA games. ~\cite{sapienza2017performance, sapienza2018non, sapienza2018individual}. 
These results pose the basis for RQ3 and RQ4: our analysis revealed that social ties can play a role in mitigating performance deterioration over the course of an activity session, however such a mitigation is not homogeneous across all individuals, but tend to benefit more the low-skilled or low-experienced ones, while high-skill or high-experience individuals may not be affected by social ties when it comes to mitigating performance deterioration.

\section{Conclusions}\label{sec:conclusions}
In this paper, we investigated four research questions. The first two questions 
focused on measuring effects within single matches: What is the influence of social ties on \textit{(RQ1)} individual players', and \textit{(RQ2)} on team's activity?. 
The other two questions focused on effects that span over the course of a gaming session (i.e., a nearly-uninterrupted sequence of consecutive matches): 
What is the influence of social ties on \textit{(RQ3)} individuals and \textit{(RQ4)} teams, over the course of entire gaming sessions?.

We set ourselves to test whether the presence of social ties affected the activity of individuals within a team (RQ1), and of the team as a whole (RQ2). We also investigated whether there exists a spillover effect by which individuals who do not play with friends, yet play in a team where some players are friends among each other, may indirectly benefit of the presence of social ties in their team. We further investigated teams composed by high/low experienced players, and high/low performing players. 
Our research revealed that individuals playing with friends have higher kills and assists frequency than players gaming without friends. Moreover, death differences suggest that players with teammates friends may have more aggressive or impulsive behaviors than those without preexisting friendships.
Moreover, kills, deaths, and assists actions are found to increase for all four categories of teams. However, for teams with preexisting ties and consisting of highly-skilled players, KDA performance dropped compared to teams with no friends. Conversely, for teams with preexisting ties but consisting of lowly-skilled players, all actions and KDA performance improve drastically compared to teams with no friends. As a summary, low-performance players (regardless of their experience) benefit the most, while high-skill players exhibit significant performance decreases, by the presence of team social ties. In other words, playing with friends benefits almost exclusively low-performance players and negatively affects high-skill players.

We also studied whether playing gaming sessions within teams with preexisting social ties affects individuals' (RQ3) and teams' (RQ4) performance over the sessions course. 
For players who only play with friends throughout a gaming session, there exists evident performance deterioration, in line with recent results~\cite{sapienza2017performance, sapienza2018non, sapienza2018individual}. However, for players that only play with strangers in a game session, results are not fully conclusive that such performance deterioration occurs.
As time goes by, teams under with social ties have gradually increasing actions and performance while teams of strangers have decreasing actions and performance. The results suggest that, except for experience, social relationships within teams can help players mitigate performance deterioration.

% Our research results have the potential to help us realize (i) how to harvest, curating and leverage longitudinal data offered by virtual teams and online games, (ii) the scientific theories that explain the impact of the interdisciplinary concept of social ties and (iii) how to untangle the particular influence of likewise motivations' impact. 
Analogous analyses along the lines of what we proposed in this paper applied to other techno-social systems only requires data that capture who interacts with whom at what point in time (or in what order). Preferably (yet not necessarily), researchers would also collect some performance/outcome data, to test whether certain interaction patterns are associated with differential levels of performance (of groups, of the individuals in the groups, or of systems of groups). Since there is often no sound theoretical argument as to exactly when or for how long an outcome is expected to occur, ideal data would include sufficiently long longitudinal observations. 
We plan to carry out some of such studies ourselves in the future, targeting other types of games, virtual teams in online and offline task-specific settings, virtual reality individual- and team-based settings, and more in general both competitive and collaborative endeavors, to study how social networks dynamics may affect human behavior and performance.

\section*{Acknowledgements}
% \bigskip Removed for double-blind review \bigskip
The authors are grateful to DARPA for support (grant \#D16AP00115). This project does not necessarily reflect the position/policy of the Government; no official endorsement should be inferred. Approved for public release; unlimited distribution.

\section*{Author contributions}
% \bigskip Removed for double-blind review \bigskip
All authors designed the research project. YZ collected the data. YZ and AS performed the experiments. All authors analyzed the results, wrote and reviewed the manuscript.

% \pagebreak
%%%%%%%%%% Insert bibliography here %%%%%%%%%%%%%%
% \balance
\bibliographystyle{IEEEtran}
\bibliography{friend}

\begin{thebibliography}{10}
\providecommand{\url}[1]{#1}
\csname url@samestyle\endcsname
\providecommand{\newblock}{\relax}
\providecommand{\bibinfo}[2]{#2}
\providecommand{\BIBentrySTDinterwordspacing}{\spaceskip=0pt\relax}
\providecommand{\BIBentryALTinterwordstretchfactor}{4}
\providecommand{\BIBentryALTinterwordspacing}{\spaceskip=\fontdimen2\font plus
\BIBentryALTinterwordstretchfactor\fontdimen3\font minus
  \fontdimen4\font\relax}
\providecommand{\BIBforeignlanguage}[2]{{%
\expandafter\ifx\csname l@#1\endcsname\relax
\typeout{** WARNING: IEEEtran.bst: No hyphenation pattern has been}%
\typeout{** loaded for the language `#1'. Using the pattern for}%
\typeout{** the default language instead.}%
\else
\language=\csname l@#1\endcsname
\fi
#2}}
\providecommand{\BIBdecl}{\relax}
\BIBdecl

\bibitem{granovetter1977strength}
M.~S. Granovetter, ``The strength of weak ties,'' in \emph{Social
  networks}.\hskip 1em plus 0.5em minus 0.4em\relax Elsevier, 1977, pp.
  347--367.

\bibitem{kawachi2001social}
I.~Kawachi and L.~F. Berkman, ``Social ties and mental health,'' \emph{Journal
  of Urban health}, vol.~78, no.~3, pp. 458--467, 2001.

\bibitem{cohen2004social}
S.~Cohen, ``Social relationships and health.'' \emph{American psychologist},
  vol.~59, no.~8, p. 676, 2004.

\bibitem{umberson2010social}
D.~Umberson and J.~Karas~Montez, ``Social relationships and health: A
  flashpoint for health policy,'' \emph{Journal of health and social behavior},
  vol.~51, no. 1\_suppl, pp. S54--S66, 2010.

\bibitem{thoits2011mechanisms}
P.~A. Thoits, ``Mechanisms linking social ties and support to physical and
  mental health,'' \emph{Journal of health \& social behavior}, vol.~52, no.~2,
  pp. 145--161, 2011.

\bibitem{backstrom2006group}
L.~Backstrom, D.~Huttenlocher, J.~Kleinberg, and X.~Lan, ``Group formation in
  large social networks: membership, growth, and evolution,'' in
  \emph{Proceedings of the 12th ACM SIGKDD international conference on
  Knowledge discovery and data mining}.\hskip 1em plus 0.5em minus 0.4em\relax
  ACM, 2006, pp. 44--54.

\bibitem{ellison2007benefits}
N.~B. Ellison, C.~Steinfield, and C.~Lampe, ``The benefits of facebook
  “friends:” social capital and college students’ use of online social
  network sites,'' \emph{Journal of Computer-Mediated Communication}, vol.~12,
  no.~4, pp. 1143--1168, 2007.

\bibitem{szell2010multirelational}
M.~Szell, R.~Lambiotte, and S.~Thurner, ``Multirelational organization of
  large-scale social networks in an online world,'' \emph{Proceedings of the
  National Academy of Sciences}, vol. 107, no.~31, pp. 13\,636--13\,641, 2010.

\bibitem{wilson2012review}
R.~E. Wilson, S.~D. Gosling, and L.~T. Graham, ``A review of facebook research
  in the social sciences,'' \emph{Perspectives on psychological science},
  vol.~7, no.~3, pp. 203--220, 2012.

\bibitem{zhong2011effects}
Z.-J. Zhong, ``The effects of collective mmorpg (massively multiplayer online
  role-playing games) play on gamers’ online and offline social capital,''
  \emph{Computers in human behavior}, vol.~27, no.~6, pp. 2352--2363, 2011.

\bibitem{trepte2012social}
S.~Trepte, L.~Reinecke, and K.~Juechems, ``The social side of gaming: How
  playing online computer games creates online and offline social support,''
  \emph{Computers in Human Behavior}, vol.~28, no.~3, pp. 832--839, 2012.

\bibitem{leonard2008richard}
T.~C. Leonard, ``Richard h. thaler, cass r. sunstein, nudge: Improving
  decisions about health, wealth, and happiness,'' 2008.

\bibitem{wuchty2007increasing}
S.~Wuchty, B.~F. Jones, and B.~Uzzi, ``The increasing dominance of teams in
  production of knowledge,'' \emph{Science}, vol. 316, no. 5827, pp.
  1036--1039, 2007.

\bibitem{guimera2005team}
R.~Guimera, B.~Uzzi, J.~Spiro, and L.~A.~N. Amaral, ``Team assembly mechanisms
  determine collaboration network structure and team performance,''
  \emph{Science}, vol. 308, no. 5722, pp. 697--702, 2005.

\bibitem{borner2010multi}
K.~B{\"o}rner, N.~Contractor, H.~J. Falk-Krzesinski, S.~M. Fiore, K.~L. Hall,
  J.~Keyton, B.~Spring, D.~Stokols, W.~Trochim, and B.~Uzzi, ``A multi-level
  systems perspective for the science of team science,'' \emph{Science
  Translational Medicine}, vol.~2, no.~49, pp. 49cm24--49cm24, 2010.

\bibitem{contractor2013some}
N.~Contractor, ``Some assembly required: leveraging web science to understand
  and enable team assembly,'' \emph{Phil. Trans. R. Soc. A}, vol. 371, no.
  1987, p. 20120385, 2013.

\bibitem{Mukherjee2018}
S.~Mukherjee, Y.~Huang, J.~Neidhardt, B.~Uzzi, and N.~Contractor, ``{Prior
  shared success predicts victory in team competitions},'' \emph{Nature Human
  Behaviour}, 2018.

\bibitem{sifa2015large}
R.~Sifa, A.~Drachen, and C.~Bauckhage, ``Large-scale cross-game player behavior
  analysis on steam,'' \emph{Borderlands}, vol.~2, pp. 46--378, 2015.

\bibitem{hamari2011framework}
J.~Hamari and V.~Eranti, ``Framework for designing and evaluating game
  achievements.'' in \emph{Digra conference}.\hskip 1em plus 0.5em minus
  0.4em\relax Citeseer, 2011.

\bibitem{becker2012analysis}
R.~Becker, Y.~Chernihov, Y.~Shavitt, and N.~Zilberman, ``An analysis of the
  steam community network evolution,'' in \emph{Electrical \& Electronics
  Engineers in Israel (IEEEI), 2012 IEEE 27th Convention of}.\hskip 1em plus
  0.5em minus 0.4em\relax IEEE, 2012, pp. 1--5.

\bibitem{blackburn2012branded}
J.~Blackburn, R.~Simha, N.~Kourtellis, X.~Zuo, M.~Ripeanu, J.~Skvoretz, and
  A.~Iamnitchi, ``Branded with a scarlet c: cheaters in a gaming social
  network,'' in \emph{Proceedings of the 21st international conference on World
  Wide Web}.\hskip 1em plus 0.5em minus 0.4em\relax ACM, 2012, pp. 81--90.

\bibitem{sapienza2017performance}
A.~Sapienza, H.~Peng, and E.~Ferrara, ``Performance dynamics and success in
  online games,'' in \emph{2017 IEEE International Conference on Data Mining
  Workshops (ICDMW)}, 2017, pp. 902--909.

\bibitem{sapienza2018non}
A.~Sapienza, A.~Bessi, and E.~Ferrara, ``Non-negative tensor factorization for
  human behavioral pattern mining in online games,'' \emph{Information},
  vol.~9, no.~3, p.~66, 2018.

\bibitem{sapienza2018individual}
A.~Sapienza, Y.~Zeng, A.~Bessi, K.~Lerman, and E.~Ferrara, ``Individual
  performance in team-based online games,'' \emph{Royal Society Open Science},
  vol.~5, no.~6, p. 180329, 2018.

\bibitem{eagle2009inferring}
N.~Eagle, A.~S. Pentland, and D.~Lazer, ``Inferring friendship network
  structure by using mobile phone data,'' \emph{Proceedings of the national
  academy of sciences}, vol. 106, no.~36, pp. 15\,274--15\,278, 2009.

\bibitem{chung2018friends}
S.~Chung, R.~B. Lount~Jr, H.~M. Park, and E.~S. Park, ``Friends with
  performance benefits: A meta-analysis on the relationship between friendship
  and group performance,'' \emph{Personality and Social Psychology Bulletin},
  vol.~44, no.~1, pp. 63--79, 2018.

\bibitem{pillemer2018friends}
J.~Pillemer and N.~P. Rothbard, ``Friends without benefits: Understanding the
  dark sides of workplace friendship,'' \emph{Academy of Management Review},
  no.~ja, 2018.

\bibitem{wentzel1997friendships}
K.~R. Wentzel and K.~Caldwell, ``Friendships, peer acceptance, and group
  membership: Realtions to academic achievement in middle school,'' \emph{Child
  development}, vol.~68, no.~6, pp. 1198--1209, 1997.

\bibitem{zimmerman2003peer}
D.~J. Zimmerman, ``Peer effects in academic outcomes: Evidence from a natural
  experiment,'' \emph{Review of Economics and statistics}, vol.~85, no.~1, pp.
  9--23, 2003.

\bibitem{ding2007peers}
W.~Ding and S.~F. Lehrer, ``Do peers affect student achievement in china's
  secondary schools?'' \emph{The Review of Economics and Statistics}, vol.~89,
  no.~2, pp. 300--312, 2007.

\bibitem{burke2013classroom}
M.~A. Burke and T.~R. Sass, ``Classroom peer effects and student achievement,''
  \emph{Journal of Labor Economics}, vol.~31, no.~1, pp. 51--82, 2013.

\bibitem{leenders2016once}
R.~T.~A. Leenders, N.~S. Contractor, and L.~A. DeChurch, ``Once upon a time:
  Understanding team processes as relational event networks,''
  \emph{Organizational Psychology Review}, vol.~6, no.~1, pp. 92--115, 2016.

\bibitem{wax2017self}
A.~Wax, L.~A. DeChurch, and N.~S. Contractor, ``Self-organizing into winning
  teams: Understanding the mechanisms that drive successful collaborations,''
  \emph{Small Group Research}, vol.~48, no.~6, pp. 665--718, 2017.

\bibitem{singer2016evidence}
P.~Singer, E.~Ferrara, F.~Kooti, M.~Strohmaier, and K.~Lerman, ``Evidence of
  online performance deterioration in user sessions on reddit,'' \emph{PloS
  one}, vol.~11, no.~8, p. e0161636, 2016.

\bibitem{ferrara2017dynamics}
E.~Ferrara, N.~Alipourfard, K.~Burghardt, C.~Gopal, and K.~Lerman, ``Dynamics
  of content quality in collaborative knowledge production,'' in
  \emph{Proceedings of 11th AAAI International Conference on Web and Social
  Media}.\hskip 1em plus 0.5em minus 0.4em\relax AAAI, 2017, pp. 520--523.

\bibitem{kooti2016}
F.~Kooti, E.~Moro, and K.~Lerman, ``Twitter session analytics: Profiling users'
  short-term behavioral changes,'' in \emph{Proceedings of the 8th
  International Conference}.\hskip 1em plus 0.5em minus 0.4em\relax Springer,
  2016, pp. 71--86.

\bibitem{kooti2017understanding}
F.~Kooti, K.~Subbian, W.~Mason, L.~Adamic, and K.~Lerman, ``Understanding
  short-term changes in online activity sessions,'' in \emph{Proceedings of the
  26th International Conference on World Wide Web Companion}, 2017, pp.
  555--563.

\end{thebibliography}


%\begin{IEEEbiography}
%[{\includegraphics[width=1in,height=1.25in,clip,keepaspectratio]{mshell}}]{Yilei Zeng}
% \newpage
% \begin{IEEEbiographynophoto}{Yilei Zeng} 

\begin{IEEEbiography}[{\includegraphics[width=1in,height=1.25in,clip,keepaspectratio]{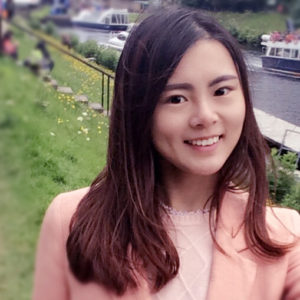}}]{\\ Yilei Zeng} 
(S'18) is a PhD student in Computer Science at the University of Southern California and a research assistant at the USC Information Sciences Institute. She holds a Master of Engineering from Peking University. Her research interests lie at the cross-road between artificial intelligence, human-computer interaction, and games.
\end{IEEEbiography}

% \begin{IEEEbiographynophoto}
\begin{IEEEbiography}[{\includegraphics[width=1in,height=1.25in,clip,keepaspectratio]{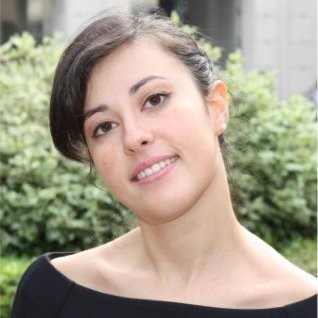}}]{\\ Anna Sapienza} 
(M'18) is a postdoctoral fellow at the USC Information Sciences Institute. She holds a PhD in Applied Mathematics from the Polytechnic University of Turin, Italy. She is interested in applying mathematical methods for the analysis of human behaviors in social networks. Her research is at the intersection between machine learning, data science,  and network science. 
\end{IEEEbiography}

%\begin{IEEEbiographynophoto}

% \begin{IEEEbiographynophoto}
\begin{IEEEbiography}[{\includegraphics[width=1in,height=1.25in,clip,keepaspectratio]{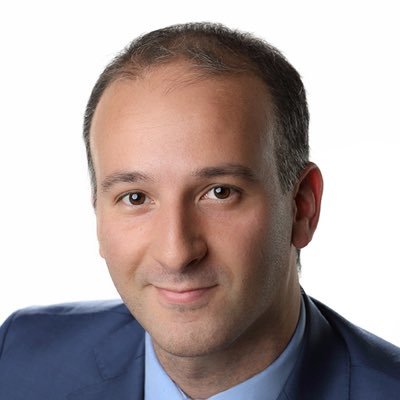}}]{\\ Emilio Ferrara} (SM'18) is Assistant Professor of Research at the University of Southern California, Research Team Leader at the USC Information Sciences Institute, and Principal Investigator at the \textit{Machine Intelligence and Data Science} (MINDS) research group. He is a recipient of the \textit{2016 DARPA Young Faculty Award} and of the \textit{2016 Complex System Society Junior Scientific Award}. His research is supported by DARPA, IARPA, AFOSR, and ONR.
\end{IEEEbiography}


% \balance
\vfill

% \enlargethispage{-5in}

\end{document}